\documentclass[twocolumn,showpacs,preprintnumbers,amsmath,amssymb]{revtex4}
\usepackage{graphicx}% Include figure files
\usepackage{dcolumn}% Align table columns on decimal point
\usepackage{bm}% bold math
\usepackage{amsmath}

\newcommand{\ba}{\begin{array}}
\newcommand{\ea}{\end{array}}
\newcommand{\be}{\begin{equation}}
\newcommand{\ee}{\end{equation}}
\begin{document}
\preprint{APS/123-QED}
\title{Lepton masses, mixings and FCNC in a minimal $S_3$-invariant\\extension of the Standard Model.}
\author{A.
  Mondrag\'on}\email{mondra@fisica.unam.mx}\affiliation{Instituto de
  F\'{\i}sica, Universidad Nacional Aut{\'o}noma de M{\'e}xico,
    Apdo. Postal 20-364,  01000 M{\'e}xico D.F., \ M{\'e}xico.}
\author{M. Mondrag\'on}\email{myriam@fisica.unam.mx}\affiliation{Instituto de    F\'{\i}sica, Universidad Nacional Aut{\'o}noma de M{\'e}xico,
    Apdo. Postal 20-364,  01000 M{\'e}xico D.F., \ M{\'e}xico.}
\author{E. Peinado} \email{eduardo@fisica.unam.mx}\affiliation{Instituto de    F\'{\i}sica, Universidad Nacional Aut{\'o}noma de M{\'e}xico,
    Apdo. Postal 20-364,  01000 M{\'e}xico D.F., \ M{\'e}xico.}
\date{\today}% It is always \today, today,
             %  but any date may be explicitly specified
\begin{abstract}
The mass matrices of the charged leptons and neutrinos, previously
derived in a minimal $S_3$-invariant extension of the Standard Model,
were reparametrized in terms of their eigenvalues. We obtained
explicit, analytical expressions for all entries
in the neutrino mixing matrix, $V_{PMNS}$, the neutrino mixing angles
and the Majorana phases as functions of the masses of charged leptons and
neutrinos in excellent agreement with the latest experimental
values. The resulting $V_{PMNS}$ matrix is very close to the tri-bimaximal
form of the neutrino mixing matrix. We also derived
explicit analytical expressions for the matrices of the Yukawa
couplings and computed the branching ratios of some selected flavour
changing neutral current processes as functions of the masses of the
charged leptons and the neutral Higgs bosons. We find that the
$S_3\times Z_2$ flavour symmetry and the strong mass hierarchy of the charged
leptons strongly suppress the FCNC processes in the leptonic sector
well below the present experimental upper bounds by many orders of magnitude.


\end{abstract}

\maketitle

\section{\label{int}Introduction}
The recent discovery of flavour oscillations of solar, atmospheric,
reactor and accelerator neutrinos have irrefutably established that
neutrinos have non-vanishing masses and mix among themselves much like
the quarks, thereby providing the first conclusive evidence of new
physics beyond the Standard
Model~\cite{jung,Mohapatra:2006gs}. Neutrino oscillation observations
and experiments, made in the past eight years, have allowed the
determination of the differences of the
neutrino masses squared and the flavour mixing angles in the leptonic
sector. The solar~\cite{altmann,smy,ahmad,aharmim},
atmospheric~\cite{fukuda,Ashie:2005ik} and
reactor~\cite{bemporad,Araki:2004mb} experiments produced the
following results:
\begin{eqnarray}
7.1\times 10^{-5}(eV)^2 \leq \Delta^2m_{12} \leq 8.9\times 10^{-5}(eV)^2,\\
\nonumber\\
0.24 \leq sin^2\theta_{12} \leq 0.40,\\
\nonumber\\
1.4\times 10^{-3}(eV)^2 \leq \Delta^2m_{13} \leq 3.3\times 10^{-3}(eV)^2,\\
\nonumber\\
0.34 \leq sin^{2}\theta_{23} \leq 0.68,
\end{eqnarray}
at $90\%$ confidence level~\cite{Maltoni:2004ei,schwetz}. For a recent
review on the phenomenology of massive neutrinos,
see~\cite{Gonzalez-Garcia:2007ib}. The CHOOZ experiment~\cite{chooz} determined an upper bound for the flavour
mixing angle between the first and the third generation:
\be
sin^{2} \theta_{13} \leq 0.046.
\ee 
However, neutrino oscillation data are insensitive to the absolute value of neutrino
masses and also to the fundamental issue of whether neutrinos are
Dirac or Majorana particles. Hence, the importance of the
upper bounds on neutrino masses provided by the searches that probe
the neutrino mass values at  rest: beta decay experiments~\cite{eitel},
neutrinoless double beta decay~\cite{eliot} and precision
cosmology ~\cite{Seljak,elgaroy,Lesgourgues}.

In the Standard Model, the Higgs and Yukawa sectors, which are
responsible for the generation of the masses of quarks and charged
leptons, do not give mass to the neutrinos. Furthermore, the Yukawa
sector of the Standard Model already has too many parameters
whose values can only be determined from experiment. These two facts
point to the necessity and convenience of extending the Standard Model
in order to make a unified and
systematic treatment of the observed hierarchies of
masses and mixings of all fermions, as well as the presence or absence of CP violating
phases in the mixing matrices. At the same time, we would also like to
reduce drastically the number of free parameters in the theory. These
two seemingly contradictory demands can be met by means of a flavour
symmetry under which the families transform in a non-trivial fashion.

Recently, we introduced a Minimal $S_{3}$-invariant Extension of the
Standard Model~\cite{kubo1} in which we argued that such a flavour
symmetry unbroken at the Fermi scale, is the permutational symmetry of
three objects $S_{3}$. In this model, we imposed $S_{3}$ as a
fundamental symmetry in the matter sector. This assumption led us
necessarily to extend the concept of flavour and generations to the
Higgs sector. Hence, going to the irreducible representations of
$S_{3}$, we added to the Higgs $SU(2)_{L}$ doublet in the
$S_{3}$-singlet representation two more Higgs $SU(2)_{L}$
doublets, which can only belong to the two components of the
$S_{3}$-doublet representation. In this way, all the matter fields in
the Minimal $S_{3}$-invariant Extension of the Standard Model - Higgs,
quark and lepton fields, including the right handed neutrino fields-
belong to the three dimensional representation ${\bf 1}\oplus{\bf 2}$
of the permutational group $S_{3}$. The leptonic sector of the model
was further constrained by an Abelian $Z_{2}$ symmetry. We found that
the $S_3 \times Z_2$ symmetry predicts the tri-bimaximal mixing and an
inverted mass hierarchy of the left handed neutrinos in good agreement
with experiment~\cite{kubo1}. More recently, we
reparametrized the mass matrices of the charged leptons and neutrinos,
previously derived in~\cite{kubo1}, in terms of their eigenvalues and
derived explicit analytical expressions for the entries in the neutrino mixing matrix, $V_{PMNS}$, and the neutrino
mixing angles and Majorana phases as functions of the masses of
charged leptons and neutrinos, in excellent agreement with the latest
experimental values~\cite{Felix:2006pn}.

The group $S_{3}$~\cite{Fritzsc1,pakvasa1,Fritzsc2,harari,Frere:1978ds,Fritzsc3,yamanaka,kaus,Fritzsch4,Harrison} and the product
groups $S_{3}\times S_{3}$~\cite{Harrison,mondragon1,mondragon2,xing} and
$S_{3}\times S_{3}\times S_{3}$~\cite{hall,hall2} have been
considered by many authors to explain successfully the hierarchical
structure of quark masses and mixings in the Standard Model. However,
in these works, the $S_{3}$, $S_{3}\times S_{3}$ and $S_{3}\times
S_{3}\times S_{3}$ symmetries are explicitly broken at the Fermi scale
to give mass to the lighter quarks and charged leptons, neutrinos are
left massless. Some other interesting models based on the $S_{3}$,
$S_{4}$, $A_{4}$ and $D_5$ flavour symmetry groups, unbroken at the Fermi
scale, have also been proposed~\cite{koide,ma,ma2,babu,chen,grimus-la,Hagedorn:2006ug,Hagedorn:2006ir}, but in
those models, equality of the number of fields and the irreducible
representations is not obtained. The generic properties of mass
textures of quarks and leptons derived in the standard model and in
supersymmetric models with a Higgs sector with non-trivial flavours and
an $S_3$ flavour symmetry have been discussed
in~\cite{Haba:2005ds,Kaneko:2006wi}. Recent flavour symmetry models are
reviewed in
~\cite{Smirnov:2006qz,Altarelli:2004za,Mondragon:2006hi,Albright:2006cw},
see also the references therein.

In this paper, we consider the flavour changing neutral current (FCNC)
processes in the Minimal $S_3$-Invariant Extension of the Standard
Model~\cite{kubo1}. After a short review of some relevant results on lepton masses
and mixings, we derive exact, explicit expressions for the matrices
of the Yukawa couplings in the leptonic sector expressed as functions
of the masses of the charged leptons and neutral Higgs bosons. With
the help of the Yukawa matrices we compute the branching ratios of
some selected FCNC processes as functions of the masses of charged
leptons and neutral Higgs bosons. We find that the interplay of the
$S_3\times Z_2$ flavour symmetry and the strong mass hierarchy of
charged leptons strongly suppresses the FCNC processes in the leptonic
sector well below the experimental upper bounds by many orders of
magnitude.
\section{The Minimal $S_{3}$-invariant Extension of the Standard
  Model}
In the Standard Model analogous fermions in different generations have
identical couplings to all gauge bosons of the strong, weak
and electromagnetic interactions. Prior to the introduction of the
Higgs boson and mass terms, the Lagrangian is chiral and invariant
with respect to permutations of the left and right fermionic fields.

The six possible permutations of three objects $(f_{1},f_{2},f_{3})$
are elements of the permutational group $S_{3}$. This is the discrete,
non-Abelian group with the smallest number of elements. The
three-dimensional real representation is not an irreducible
representation of $S_{3}$. It can be decomposed into the direct
sum of a doublet $f_{D}$ and a singlet $f_{s}$, where
%{\small
\be
\begin{array}{l}
f_{s}=\frac{1}{\sqrt{3}}(f_{1}+f_{2}+f_{3}),\\
\\
f_{D}^{T}=\left(\frac{1}{\sqrt{2}}(f_{1}-f_{2}),\frac{1}{\sqrt{6}}(f_{1}+f_{2}-2f_{3})\right).
\end{array}
\ee%}
The direct product of two doublets ${\bf p_{D}}^{T} =(p_{D1},p_{D2})$
and ${\bf q_{D}}^{T}=(q_{D1},q_{D2})$ may be decomposed into the direct
sum of two singlets ${\bf r_{s}}$ and ${\bf r_{s'}}$, and one doublet
${\bf r_{D}}^{T}$ where
%{\small
\be
\begin{array}{lr}
{\bf r_{s}} = p_{D1} q_{D1} + p_{D2}q_{D2}, & {\bf r_{s'}} =
p_{D1}q_{D2} - p_{D2}q_{D1},
\end{array}
\ee
\be
{\bf r_{D}}^{T}= (r_{D1},r_{D2})=(p_{D1} q_{D2} + p_{D2}q_{D1},p_{D1} q_{D1} - p_{D2}q_{D2}).
\ee
%}
The antisymmetric singlet ${\bf r_{s'}}$ is not invariant under $S_{3}$.

Since the Standard Model has only one Higgs $SU(2)_{L}$ doublet,
which can only be an $S_{3}$ singlet, it can only give mass to the
quark or charged lepton in the $S_{3}$ singlet representation, one in
each family, without breaking the $S_{3}$ symmetry.

Hence, in order to impose $S_{3}$ as a fundamental symmetry, unbroken
at the Fermi scale, we are led to extend the Higgs sector of the
theory. The quark, lepton and Higgs fields are
%{\small
\be
\begin{split}
Q^T=(u_L,d_L)~,~ u_R~,~d_R~,~\\L^T=(\nu_L,e_L)~,~e_R~,~ 
\nu_R~\mbox{ and }~H,
\end{split}
\ee%}
in an obvious notation. All of these fields have three species, and
we assume that each one forms a reducible representation ${\bf 1}_S\oplus{\bf 2}$.
The doublets carry capital indices $I$ and $J$, which run from $1$ to $2$,
and the singlets are denoted by
$Q_3,~u_{3R},~d_{3R},~L_3,~e_{3R},~\nu_{3R}$ and $~H_S$. Note that the subscript $3$ denotes the
singlet representation and not the third generation.
The most general renormalizable Yukawa interactions of this model are given by
\be
{\cal L}_Y = {\cal L}_{Y_D}+{\cal L}_{Y_U}
+{\cal L}_{Y_E}+{\cal L}_{Y_\nu},
\ee
where
%{\small
\be
\begin{array}{lll}
{\cal L}_{Y_D} &=&
- Y_1^d \overline{ Q}_I H_S d_{IR} - Y_3^d \overline{ Q}_3 H_S d_{3R}  \\
&  &   -Y^{d}_{2}[~ \overline{ Q}_{I} \kappa_{IJ} H_1  d_{JR}
+\overline{ Q}_{I} \eta_{IJ} H_2  d_{JR}~]\\
&  & -Y^d_{4} \overline{ Q}_3 H_I  d_{IR} - Y^d_{5} \overline{ Q}_I H_I d_{3R} 
+~\mbox{h.c.} ,
\label{lagd}
\end{array}
\ee
\be
\begin{array}{lll}
{\cal L}_{Y_U} &=&
-Y^u_1 \overline{ Q}_{I}(i \sigma_2) H_S^* u_{IR} 
-Y^u_3\overline{ Q}_3(i \sigma_2) H_S^* u_{3R} \\
&  &   -Y^{u}_{2}[~ \overline{ Q}_{I} \kappa_{IJ} (i \sigma_2)H_1^*  u_{JR}
+\eta  \overline{ Q}_{I} \eta_{IJ}(i \sigma_2) H_2^*  u_{JR}~]\\
&  &
-Y^u_{4} \overline{ Q}_{3} (i \sigma_2)H_I^* u_{IR} 
-Y^u_{5}\overline{ Q}_I (i \sigma_2)H_I^*  u_{3R} +~\mbox{h.c.},
\label{lagu}
\end{array}
\ee
\be
\begin{array}{lll}
{\cal L}_{Y_E} &=& -Y^e_1\overline{ L}_I H_S e_{IR} 
-Y^e_3 \overline{ L}_3 H_S e_{3R} \\
&  &  - Y^{e}_{2}[~ \overline{ L}_{I}\kappa_{IJ}H_1  e_{JR}
+\overline{ L}_{I} \eta_{IJ} H_2  e_{JR}~]\\
 &  & -Y^e_{4}\overline{ L}_3 H_I e_{IR} 
-Y^e_{5} \overline{ L}_I H_I e_{3R} +~\mbox{h.c.},
\end{array}
\label{lage}
\ee
\be
\begin{array}{lcl}
{\cal L}_{Y_\nu} &=& -Y^{\nu}_1\overline{ L}_I (i \sigma_2)H_S^* \nu_{IR} 
-Y^\nu_3 \overline{ L}_3(i \sigma_2) H_S^* \nu_{3R} \\
&  &   -Y^{\nu}_{2}[~\overline{ L}_{I}\kappa_{IJ}(i \sigma_2)H_1^*  \nu_{JR}
+ \overline{ L}_{I} \eta_{IJ}(i \sigma_2) H_2^*  \nu_{JR}~]\\
 &  & -Y^\nu_{4}\overline{ L}_3(i \sigma_2) H_I^* \nu_{IR} 
-Y^\nu_{5} \overline{ L}_I (i \sigma_2)H_I^* \nu_{3R}+~\mbox{h.c.},
\label{lagnu}
\end{array}
\ee%}
and
\be
\kappa = \left( \begin{array}{cc}
0& 1\\ 1 & 0\\
\end{array}\right)~~\mbox{and}~~
\eta = \left( \begin{array}{cc}
1& 0\\ 0 & -1\\
\end{array}\right).
\label{kappa}
\ee Furthermore, we add to the Lagrangian the Majorana mass terms for
the right-handed neutrinos \be {\cal L}_{M} = -M_1 \nu_{IR}^T C
\nu_{IR} -M_3 \nu_{3R}^T C \nu_{3R}.
\label{majo}
\ee

Due to the presence of three Higgs fields, the Higgs potential
$V_H(H_S,H_D)$ is more complicated than that of the Standard
Model. This potential was analyzed by Pakvasa and Sugawara~\cite{pakvasa1} who
found that in addition to the $S_{3}$ symmetry, it has a permutational
symmetry
$S_{2}^{\prime}$: $H_{1}\leftrightarrow H_{2}$, which is not a subgroup of the
flavour group $S_{3}$. In this communication, we will assume that the
vacuum respects the accidental $S_{2}^{\prime}$ symmetry of the Higgs potential
and that
\be
\langle H_{1} \rangle = \langle H_{2} \rangle.
\ee

With these assumptions, the Yukawa interactions, eqs. (\ref{lagd})-(\ref{lagnu}) yield mass matrices,
for all fermions in the theory, of the general form~\cite{kubo1}
\be
{\bf M} = \left( \begin{array}{ccc}
\mu_{1}+\mu_{2} & \mu_{2} & \mu_{5} 
\\  \mu_{2} & \mu_{1}-\mu_{2} &\mu_{5}
  \\ \mu_{4} & \mu_{4}&  \mu_{3}
\end{array}\right).
\label{general-m}
\ee
The Majorana mass for the left handed neutrinos $\nu_{L}$ is generated
by the see-saw mechanism. The corresponding mass matrix is
given by
\be
{\bf M_{\nu}} = {\bf M_{\nu_D}}\tilde{{\bf M}}^{-1}({\bf M_{\nu_D}})^T,
\label{seesaw}
\ee
where $\tilde{{\bf M}}=\mbox{diag}(M_1,M_1,M_3)$.
\\
In principle, all entries in the mass matrices can be complex since
there is no restriction coming from the flavour symmetry $S_{3}$.
The mass matrices are diagonalized by bi-unitary transformations as
%{\small
\be
\begin{array}{rcl}
U_{d(u,e)L}^{\dag}{\bf M}_{d(u,e)}U_{d(u,e)R} 
&=&\mbox{diag} (m_{d(u,e)}, m_{s(c,\mu)},m_{b(t,\tau)}),
\\ 
\\
U_{\nu}^{T}{\bf M_\nu}U_{\nu} &=&
\mbox{diag} (m_{\nu_1},m_{\nu_2},m_{\nu_3}).
\end{array}
\label{unu}
\ee
%}
The entries in the diagonal matrices may be complex, so the physical
masses are their absolute values.

The mixing matrices are, by definition,
\be
\begin{array}{ll}
V_{CKM} = U_{uL}^{\dag} U_{dL},& V_{PMNS} = U_{eL}^{\dag} U_{\nu} K,
\label{ckm1}
\end{array}
\ee
where $K$ is the diagonal matrix of the Majorana phase factors.
\section{The mass matrices in the leptonic sector and $Z_{2}$
  symmetry}
A further reduction of the number of parameters in the leptonic sector
may be achieved by means of an Abelian $Z_{2}$ symmetry. A possible set
of charge assignments of $Z_{2}$, compatible with the experimental
data on masses and mixings in the leptonic sector is given in Table~\ref{table1}.

%\begin{center}
%\normalsize
\begin{table}
\caption{\label{table1}$Z_2$ assignment in the leptonic sector.}
\begin{ruledtabular}
\begin{tabular}{cc}
\hline
 $-$ &  $+$
\\ %\hline

$H_S, ~\nu_{3R}$ & $H_I, ~L_3, ~L_I, ~e_{3R},~ e_{IR},~\nu_{IR}$
\\ %\hline
\end{tabular}
\end{ruledtabular}
\end{table}
%\end{center}
These $Z_2$ assignments forbid the following Yukawa couplings
\be
 Y^e_{1} = Y^e_{3}= Y^{\nu}_{1}= Y^{\nu}_{5}=0.
\label{zeros}
\ee
Therefore, the corresponding entries in the mass matrices vanish, {\it
  i.e.}, $\mu_{1}^{e}=\mu_{3}^{e}=0$ and $\mu_{1}^{\nu}=\mu_{5}^{\nu}=0$.
\begin{center}{\it The mass matrix of the charged leptons}\end{center}
The mass matrix of the charged leptons takes the form
\be
M_{e} = m_{\tau}\left( \begin{array}{ccc}
\tilde{\mu}_{2} & \tilde{\mu}_{2} & \tilde{\mu}_{5} 
\\  \tilde{\mu}_{2} &-\tilde{\mu}_{2} &\tilde{\mu}_{5}
  \\ \tilde{\mu}_{4} & \tilde{\mu}_{4}& 0
\end{array}\right).
\label{charged-leptons-m}
\ee
The unitary matrix $U_{eL}$ that enters in the definition of the
mixing matrix, $V_{PMNS}$, is calculated from
\be
U_{eL}^{\dag}M_{e}M_{e}^{\dag}U_{eL}=\mbox{diag}(m_{e}^{2},m_{\mu}^{2},m_{\tau}^{2}),
\ee
where $m_{e}$, $m_{\mu}$ and $m_{\tau}$ are the masses of the charged
leptons, and
%\begin{widetext}
\begin{equation}
\frac{M_{e}M_{e}^{\dag}}{m_{\tau}^2}= \left( \begin {array}{ccc} 2|\tilde{\mu}_{2}|^2+|\tilde{\mu}_{5}|^{2}&|\tilde{\mu}_{5}|^{2}&2|\tilde{\mu}_{2}||\tilde{\mu}_{4}|e^{-i\delta_{e}}\\\noalign{\medskip}|\tilde{\mu}_{5}|^{2}&2|\tilde{\mu}_{2}|^{2}+|\tilde{\mu}_{5}|^{2}&0\\\noalign{\medskip}2|\tilde{\mu}_{2}||\tilde{\mu}_{4}|e^{i\delta_{e}}&0&2\,|\tilde{\mu}_{4}|^{2}\end {array} \right).
\label{mmdag}
\end{equation}
%\end{widetext}
%since the matrix $M_{e}M_{e}^{\dag}$ has only one complex entry,
%$\tilde{\mu}_{2}\tilde{\mu}_{4}$, we may take $\tilde{\mu}_{2}$ real.
Notice that this matrix has only one non-ignorable phase factor.

The parameters $|\tilde{\mu}_{2}|$, $|\tilde{\mu}_{4}|$ and
$|\tilde{\mu}_{5}|$ may readily be expressed in terms of the charged
lepton masses. From the invariants of $M_{e}M_{e}^{\dag}$, we get the
set of equations
\be
%\begin{array}{l}
\begin{split}
 Tr(M_{e}M_{e}^{\dag})=m_{e}^{2}+m_{\mu}^{2}+m_{\tau}^{2}
=\\
\\m_{\tau}^{2}\left[4|\tilde{\mu}_{2}|^{2}+2\left(|\tilde{\mu}_{4}|^{2}+|\tilde{\mu}_{5}|^{2}\right)
  \right],
%\end{array}
\end{split}
\label{trace}
\ee
\be
\begin{split}
%\begin{array}{lcl}
\chi(M_{e}M_{e}^{\dag})=m_{\tau}^{2}(m_{e}^{2}+m_{\mu}^{2})+m_{e}^{2}m_{\mu}^{2}=\\\\
4m_{\tau}^{4}\left[
|\tilde{\mu}_{2}|^{4}+|\tilde{\mu}_{2}|^{2}\left(|\tilde{\mu}_{4}|^{2}+|\tilde{\mu}_{5}|^{2}\right)+|\tilde{\mu}_{4}|^{2}|\tilde{\mu}_{5}|^{2}
\right],
\end{split}
\label{chis}
\ee
\begin{flushleft}
\be
\begin{array}{l}
det(M_{e}M_{e}^{\dag})=m_{e}^{2}m_{\mu}^{2}m_{\tau}^{2}
=4m_{\tau}^{6}|\tilde{\mu}_{2}|^{2}|\tilde{\mu}_{4}|^{2}|\tilde{\mu}_{5}|^{2},
\end{array}
\label{determinant}
\ee
\end{flushleft}
%}
where {\small
$\chi(M_{e}M_{e}^{\dag})=\frac{1}{2}\left[(Tr(M_{e}M_{e}^{\dag}))^{2}-Tr(M_{e}M_{e}^{\dag})^2
\right]$}.

Solving these equations for $|\tilde{\mu}_{2}|^2$, $|\tilde{\mu}_{4}|^2$ and
$|\tilde{\mu}_{5}|^2$, we obtain
\be
|\tilde{\mu}_{2}|^2=\frac{\tilde{m}_{\mu}^2}{2}\frac{1+x^4}{1+x^2}+\beta,
\label{mu2beta}
\ee
and 
\begin{widetext}
\be
\begin{split}
\begin{array}{l}
|\tilde{\mu}_{4,5}|^2=
\frac{1}{4}\left(1-\tilde{m}_{\mu}^2\frac{(1-x^2)^2}{1+x^2}-4\beta
\right)\\ \\
\mp\frac{1}{4}\left[\left(1-\tilde{m}_{\mu}^2\frac{(1-x^2)^2}{1+x^2}\right)^2-8\tilde{m}_{e}^2\frac{1+x^2}{1+x^4}+8\beta\left(1-\tilde{m}_{\mu}^2\frac{(1-x^2)^2}{1+x^2}+
\frac{x^2}{1+\frac{2\beta(1+x^2) }{\tilde{m}_{\mu}^2(1+x^4)}}\frac{(1+x^2)^2}{(1+x^4)^2}\right)+16\beta^2\right]^{1/2}.
\end{array}
\end{split}
\ee
\end{widetext}
In these expressions, $x=m_e/m_\mu$, $\tilde{m}_{\mu}=m_{\mu}/m_{\tau}$ and $\beta$ is the smallest solution of the equation
%{\small
\begin{equation}
\begin{array}{l}
\beta^3-\frac{1}{2}(1-2y+6\frac{z}{y})\beta^2-\frac{1}{4}(y-y^{2}-4\frac{z}{y}+7z-12\frac{z^{2}}{y^{2}})\beta-
\\ \\
\frac{1}{8}yz-\frac{1}{2}\frac{z^{2}}{y^{2}}+\frac{3}{4}\frac{z^{2}}{y}-\frac{z^{3}}{y^{3}}=0,
\label{cubic4beta}
\end{array}
\end{equation}
%}
where $y=(m_{e}^{2}+m_{\mu}^2)/m_{\tau}^{2}$ and $z=m_{\mu}^2m_{e}^{2}/m_{\tau}^{4}$.\\
A good, order of magnitude, estimate for $\beta$ is obtained from
(\ref{cubic4beta})
\be
\beta \approx
-\frac{m_{\mu}^2m_{e}^{2}}{2m_{\tau}^{2}(m_{\tau}^{2}-(m_{\mu}^2+m_{e}^{2}))}.
\ee
Once $M_{e}M_{e}^{\dag}$ has been reparametrized in terms of the
charged lepton masses, it is straightforward to compute $M_e$ and $U_{eL}$ also
as functions of the charged lepton masses~\cite{Felix:2006pn}. 
The resulting expression for $M_e$, written to order
$\left(m_{\mu}m_{e}/m_{\tau}^{2}\right)^{2}$ and
$x^{4}=\left(m_{e}/m_{\mu}\right)^4$ is
\begin{widetext}
\be
M_{e}\approx m_{\tau} \left( 
\begin{array}{ccc}
\frac{1}{\sqrt{2}}\frac{\tilde{m}_{\mu}}{\sqrt{1+x^2}} & \frac{1}{\sqrt{2}}\frac{\tilde{m}_{\mu}}{\sqrt{1+x^2}} & \frac{1}{\sqrt{2}} \sqrt{\frac{1+x^2-\tilde{m}_{\mu}^2}{1+x^2}} \\ \\
 \frac{1}{\sqrt{2}}\frac{\tilde{m}_{\mu}}{\sqrt{1+x^2}} &-\frac{1}{\sqrt{2}}\frac{\tilde{m}_{\mu}}{\sqrt{1+x^2}}  & \frac{1}{\sqrt{2}} \sqrt{\frac{1+x^2-\tilde{m}_{\mu}^2}{1+x^2}} \\ \\
\frac{\tilde{m}_{e}(1+x^2)}{\sqrt{1+x^2-\tilde{m}_{\mu}^2}}e^{i\delta_{e}} & \frac{\tilde{m}_{e}(1+x^2)}{\sqrt{1+x^2-\tilde{m}_{\mu}^2}}e^{i\delta_{e}} & 0
\end{array}
\right).
\label{emass}
\ee
\end{widetext}
This approximation is numerically exact up to order $10^{-9}$ in units
of the $\tau$ mass. Notice that this matrix has no free parameters
other than the Dirac phase $\delta_e$.

The unitary matrix $U_{eL}$ that diagonalizes $M_{e}M_{e}^{\dagger}$ and
enters in the definition of the neutrino mixing matrix $V_{PMNS}$ may
be written as
\be
\ba{l}
U_{eL}= \left(\ba{ccc} 
1& 0 & 0 \\
0 & 1 & 0 \\
0 & 0 & e^{i\delta_{e}}
\ea\right) \left(
\ba{ccc}
O_{11}& -O_{12}& O_{13} \\
-O_{21}& O_{22}& O_{23} \\
-O_{31}& -O_{32}& O_{33} 
\ea
\right)~,
\ea
\label{unitary-leptons}
\ee
where the orthogonal matrix ${\bf O}_{eL}$ in the right hand side of
eq. (\ref{unitary-leptons}), written to the same order of magnitude as
$M_e$, is
\begin{widetext}
\be
%\begin{split}
%\begin{array}{l}
{\bf O}_{eL}\approx%\\\\
\left(
\ba{ccc}
\frac{1}{\sqrt{2}}x
\frac{(
1+2\tilde{m}_{\mu}^2+4x^2+\tilde{m}_{\mu}^4+2\tilde{m}_{e}^2
)}{\sqrt{1+\tilde{m}_{\mu}^2+5x^2-\tilde{m}_{\mu}^4-\tilde{m}_{\mu}^6+\tilde{m}_{e}^2+12x^4}}&
-\frac{1}{\sqrt{2}}\frac{(1-2\tilde{m}_{\mu}^2+\tilde{m}_{\mu}^4-2\tilde{m}_{e}^2)}{\sqrt{1-4\tilde{m}_{\mu}^2+x^2+6\tilde{m}_{\mu}^4-4\tilde{m}_{\mu}^6-5\tilde{m}_{e}^2}}
& \frac{1}{\sqrt{2}} \\ \\
-\frac{1}{\sqrt{2}}x
\frac{(
1+4x^2-\tilde{m}_{\mu}^4-2\tilde{m}_{e}^2
)}{\sqrt{1+\tilde{m}_{\mu}^2+5x^2-\tilde{m}_{\mu}^4-\tilde{m}_{\mu}^6+\tilde{m}_{e}^2+12x^4}}
&
\frac{1}{\sqrt{2}}\frac{(1-2\tilde{m}_{\mu}^2+\tilde{m}_{\mu}^4)}{\sqrt{1-4\tilde{m}_{\mu}^2+x^2+6\tilde{m}_{\mu}^4-4\tilde{m}_{\mu}^6-5\tilde{m}_{e}^2}}
& \frac{1}{\sqrt{2}} \\ \\
-\frac{\sqrt{1+2x^2-\tilde{m}_{\mu}^2-\tilde{m}_{e}^2}(1+\tilde{m}_{\mu}^2+x^2-2\tilde{m}_{e}^2)}{\sqrt{1+\tilde{m}_{\mu}^2+5x^2-\tilde{m}_{\mu}^4-\tilde{m}_{\mu}^6+\tilde{m}_{e}^2+12x^4}} & -x\frac{(1+x^2-\tilde{m}_{\mu}^2-2\tilde{m}_{e}^2)\sqrt{1+2x^2-\tilde{m}_{\mu}^2-\tilde{m}_{e}^2}}{\sqrt{1-4\tilde{m}_{\mu}^2+x^2+6\tilde{m}_{\mu}^4-4\tilde{m}_{\mu}^6-5\tilde{m}_{e}^2}} &\frac{\sqrt{1+x^2}\tilde{m}_{e}\tilde{m}_{\mu}}{\sqrt{1+x^2-\tilde{m}_{\mu}^2}}
\ea
\right)~,
\label{unitary-leptons-2}
%\end{array}
%\end{split}
\ee
\end{widetext}
where, as before, $\tilde{m_{\mu}}=m_{\mu}/m_{\tau}$,
$\tilde{m_{e}}=m_{e}/m_{\tau}$ and $x=m_{e}/m_{\mu}$.
%\newpage
\begin{center}{\it The mass matrix of the neutrinos}\end{center}
According to the $Z_{2}$ selection rule eq. (\ref{zeros}), the mass
matrix of the Dirac neutrinos takes the form
\be
{\bf M_{\nu_D}} = \left( \begin{array}{ccc}
\mu^{\nu}_{2} & \mu^{\nu}_{2} & 0
\\  \mu^{\nu}_{2} & -\mu^{\nu}_{2} &0
  \\ \mu^{\nu}_{4} & \mu^{\nu}_{4}&  \mu^{\nu}_{3}
\end{array}\right).
\label{neutrinod-m}
\ee
\\
Then, the mass matrix for the left-handed Majorana neutrinos, ${\bf
  M_{\nu}}$, obtained
from the see-saw mechanism, ${\bf M_{\nu}} = {\bf M_{\nu_D}}\tilde{{\bf M}}^{-1} 
({\bf M_{\nu_D}})^T$, is
\be
%\begin{array}{l}
{\bf M_{\nu}} = %{\bf M_{\nu_D}}\tilde{{\bf M}}^{-1} 
%({\bf M_{\nu_D}})^T=%\\
%\\
\left( \begin{array}{ccc}
2 (\rho^{\nu}_{2})^2 & 0 & 
2 \rho^{\nu}_2 \rho^{\nu}_{4}
\\ 0 & 2 (\rho^{\nu}_{2})^2 & 0
  \\ 2 \rho^{\nu}_2 \rho^{\nu}_{4} & 0  &  
2 (\rho^{\nu}_{4})^2 +
(\rho^{\nu}_3)^2
\end{array}\right),
%\end{array}
\label{m-nu}
\ee
where $\rho_2^\nu =(\mu^{\nu}_2)/M_{1}^{1/2}$,  $\rho_4^\nu
=(\mu^{\nu}_4)/M_{1}^{1/2}$ and $\rho_3^\nu
=(\mu^{\nu}_3)/M_{3}^{1/2}$; $M_{1}$ and $M_{3}$ are the masses of
the right handed neutrinos appearing in (\ref{majo}).

The non-Hermitian, complex, symmetric neutrino mass matrix $M_{\nu}$ may be brought
to a diagonal form by a bi-unitary transformation, as
\be
U_{\nu}^{T}M_{\nu}U_{\nu}=\mbox{diag}\left(|m_{\nu_{1}}|e^{i\phi_{1}},|m_{\nu_{2}}|e^{i\phi_{2}},|m_{\nu_{3}}|e^{i\phi_{\nu}}\right),
\label{diagneutrino}
\ee
where $U_{\nu}$ is the matrix that diagonalizes the matrix
$M_{\nu}^{\dagger}M_{\nu}$.\\
In order to compute $U_{\nu}$, we notice that $M_{\nu}^{\dagger}M_{\nu}$ has the same
texture zeroes as $M_{\nu}$
\be
M_{\nu}^{\dagger}M_{\nu}=\left(
\ba{ccc}
|A|^2+ |B|^2 & 0 & A^{\star}B+B^{\star}D \\
0 & |A|^2 & 0 \\
AB^{\star}+BD^{\star} & 0 &  |B|^2+|D|^2
\ea\right),
\ee
where $A=2 (\rho^{\nu}_{2})^2$, $B=2 \rho^{\nu}_2 \rho^{\nu}_{4}$, and $D=2 (\rho^{\nu}_{4})^2 +
(\rho^{\nu}_3)^2$.\\
Furthermore, notice that the entries in the upper right corner and lower left
corner are complex conjugates of each other, all other entries are
real. Therefore, the matrix $U_{\nu L}$ that diagonalizes
$M_{\nu}^{\dagger}M_{\nu}$, takes the form
\be
U_{\nu}=\left(\ba{ccc} 
1& 0 & 0 \\
0 & 1 & 0 \\
0 & 0 & e^{i\delta_{\nu}} 
\ea\right)\left(
\begin{array}{ccc}
\cos \eta & \sin \eta & 0 \\
0 & 0 & 1 \\
-\sin \eta  & \cos \eta &0
\end{array}
\right).
\label{ununew}
\ee
If we require that the defining equation (\ref{diagneutrino}) be
satisfied as an identity, we get the following set of equations:
\be
\ba{l}
2 (\rho^{\nu}_{2})^2=m_{\nu_{3}},\\
\\
2 (\rho^{\nu}_{2})^2=m_{\nu_{1}}\cos^2 \eta + m_{\nu_{2}}\sin^2 \eta, \\
\\
2 \rho^{\nu}_2 \rho^{\nu}_{4}=\sin \eta \cos \eta (m_{\nu_{2}}-m_{\nu_{1}})e^{-i\delta_{\nu}},\\
\\
2 (\rho^{\nu}_{4})^2 +
(\rho^{\nu}_3)^2=(m_{\nu_{1}}\sin^2 \eta + m_{\nu_{2}}\cos^2 \eta)e^{-2i\delta_{\nu}}.
\ea
\ee
Solving these equations for $\sin \eta$ and $\cos \eta$, we find
\be
\ba{lr}
\sin^2\eta=\frac{m_{\nu_{3}}-m_{\nu_{1}}}{m_{\nu_{2}}-m_{\nu_{1}}},
&
\cos^2\eta=\frac{m_{\nu_{2}}-m_{\nu_{3}}}{m_{\nu_{2}}-m_{\nu_{1}}}.
\ea
\label{cosysin}
\ee
Hence, the matrices $M_{\nu}$ and $U_\nu$, reparametrized in terms of the
complex neutrino masses, take the form~\cite{Felix:2006pn}
\begin{widetext}
\be
M_{\nu} = 
\left( \begin{array}{ccc}
m_{\nu_{3}} & 0 & \sqrt{(m_{\nu_{3}}-m_{\nu_{1}})(m_{\nu_{2}}-m_{\nu_{3}})}e^{-i\delta_{\nu}}
\\ 
0 &m_{\nu_{3}}  & 0
\\
\sqrt{(m_{\nu_{3}}-m_{\nu_{1}})(m_{\nu_{2}}-m_{\nu_{3}})} e^{-i\delta_{\nu}} & 0  & (m_{\nu_{1}}+m_{\nu_{2}}-m_{\nu_{3}})e^{-2i\delta_{\nu}}
\end{array}\right).
\label{m-nu2}
\ee
\end{widetext}
and
\be
U_{\nu}=
\left(\ba{ccc} 
1& 0 & 0 \\
0 & 1 & 0 \\
0 & 0 & e^{i\delta_{\nu}} 
\ea\right)\left(
\begin{array}{ccc}
\sqrt{\displaystyle{\frac{m_{\nu_{2}}-m_{\nu_{3}}}
{m_{\nu_{2}}-m_{\nu_{1}}}}}& 
\sqrt{
\displaystyle{\frac{m_{\nu_{3}}-m_{\nu_{1}}}{m_{\nu_{2}}-m_{\nu_{1}}}}} & 0\\
0&0&1\\
-\sqrt{
\displaystyle{\frac{m_{\nu_{3}}-m_{\nu_{1}}}{m_{\nu_{2}}-m_{\nu_{1}}}}}  &\sqrt{\displaystyle{\frac{m_{\nu_{2}}-m_{\nu_{3}}}
{m_{\nu_{2}}-m_{\nu_{1}}}}}&0
\end{array}
\right).
\label{unu-final}
\ee
The unitarity of $U_{\nu}$ constrains $\sin \eta$ to be real and thus 
$|\sin \eta|\leq 1$, this condition fixes the phases $\phi_{1}$ and
$\phi_{2}$ as
\be
|m_{\nu_{1}}|\sin \phi_{1}=|m_{\nu_{2}}|\sin \phi_{2}=|m_{\nu_{3}}|\sin \phi_{\nu}.
\label{phase-condition}
\ee
The only free parameters in these matrices, are the phase $\phi_{\nu}$, implicit in $m_{\nu_{1}}$,
$m_{\nu_{2}}$ and $m_{\nu_{3}}$, and the Dirac phase $\delta_{\nu}$.
\bigskip
\begin{center}
{\it The neutrino mixing matrix}
\end{center}
The neutrino mixing matrix $V_{PMNS}$, is the product
$U_{eL}^{\dagger}U_{\nu}K$, where $K$ is the diagonal matrix of the
Majorana phase factors, defined by
\be
diag(m_{\nu_{1}},m_{\nu_{2}},m_{\nu_{3}})=K^{\dagger}diag(|m_{\nu_{1}}|,|m_{\nu_{2}}|,|m_{\nu_{3}}|)K^{\dagger}.
\ee
Except for an overall phase factor $e^{i\phi_{1}}$, which can be
ignored, $K$ is 
\be
K=diag(1,e^{i\alpha},e^{i\beta}),
\ee
where $\alpha=1/2(\phi_{1}-\phi_{2})$ and
$\beta=1/2(\phi_{1}-\phi_{\nu})$ are the Majorana phases. 

Therefore, the theoretical mixing matrix $V_{PMNS}^{th}$, is given by
\begin{widetext}
\be
\begin{split}
V_{PMNS}^{th}=
\left(
\ba{ccc}
O_{11}\cos \eta + O_{31}\sin \eta e^{i\delta} & O_{11}\sin
\eta-O_{31} \cos \eta e^{i\delta} & -O_{21}  \\ \\
-O_{12}\cos \eta + O_{32}\sin \eta e^{i\delta} & -O_{12}\sin
\eta-O_{32}\cos \eta e^{i\delta} & O_{22} \\ \\
O_{13}\cos \eta - O_{33}\sin \eta e^{i\delta} & O_{13}\sin
\eta+O_{33}\cos \eta e^{i\delta} & O_{23} 
\ea
\right)%\\
 \times K,
\end{split}
\label{vpmns2}
\ee
\end{widetext}
where $\cos \eta$ and $\sin \eta$ are given in eq (\ref{cosysin}),
$O_{ij}$ are given in eq (\ref{unitary-leptons}) and
(\ref{unitary-leptons-2}), and $\delta=\delta_{\nu}-\delta_{e}$. 

To find the relation of our results
with the neutrino mixing angles we make use of the equality of the
absolute values of the elements of $V_{PMNS}^{th}$ and
$V_{PMNS}^{PDG}$~\cite{PDG}, that is
\be
|V_{PMNS}^{th}|=|V_{PMNS}^{PDG}|.
\label{mix-cond}
\ee
This relation allows us to derive expressions for the mixing angles
in terms of the charged lepton and neutrino masses 
\be
\ba{ll}
%\begin{split}
|\sin \theta_{13}|=O_{21}
, &
%\\
|\sin \theta_{23}|= \frac{O_{22}}{\sqrt{O_{22}^2+O_{23}^2}}
\ea
\label{tan}
\ee
and
\be
 |\tan \theta_{12}|^2= \cot^2 \eta \frac{O_{11}^2\frac{1}{\cot^2
    \eta}+O_{31}^2-2O_{31}O_{11}\frac{1}{\cot \eta} \cos \delta}{O_{11}^2\cot^2 \eta+O_{31}^2+2O_{31}O_{11}\cot \eta \cos \delta}.
\label{tan2}
\ee
The magnitudes of the reactor and atmospheric mixing angles,
$\theta_{13}$ and $\theta_{23}$, are determined by the masses of the
charged leptons only. Keeping terms up to order $(m_{e}^2/m_{\mu}^2)$ and
$(m_{\mu}/m_{\tau})^4$, we get
%{\small
\be
%\ba{lr}
\begin{split}
\sin \theta_{13}\approx \frac{1}{\sqrt{2}}x
\frac{(
1+4x^2-\tilde{m}_{\mu}^4)}{\sqrt{1+\tilde{m}_{\mu}^2+5x^2-\tilde{m}_{\mu}^4}}
,% &
\\ \\
\sin \theta_{23}\approx  \frac{1}{\sqrt{2}}\frac{1-2\tilde{m}_{\mu}^2+\tilde{m}_{\mu}^4}{\sqrt{1-4\tilde{m}_{\mu}^2+x^2+6\tilde{m}_{\mu}^4}}.
%\ea
\end{split}
\label{mixing-angles}
\ee
Substitution of the small numerical values
$\tilde{m}_{\mu}=5.94\times 10^{-2}$ and $x=m_{e}/m_{\mu}=4.84 \times 10^{-3}$$\tilde{m}_{\mu}=m_{\mu}/m_{\tau}=5.95\times 10^{-2}$ for the leptonic mass ratios
$\tilde{m}_{\mu}$ and $x$ in the right hand side of
(\ref{mixing-angles}) yields the numerical values of $\sin
\theta_{13}$ and $\sin \theta_{23}$
\be
\ba{l}
\sin \theta_{13}=0.0034 \\ \\
\sin \theta_{23} =\frac{1}{\sqrt{2}}-8.4\times 10^{-6}.
\ea
\ee
From these numbers, it is evident that the theoretical values of $\sin
\theta_{13}$ and $\sin \theta_{23}$ are very close to the
corresponding tri-bimaximal mixing values $\sin \theta_{13}^{tri}=0$
and $\sin \theta_{23}^{tri}=1/\sqrt{2}$~\cite{Harrison:2002er}.

The dependence of $\tan \theta_{12}$ on the Dirac phase $\delta$, see
(\ref{tan2}), is very weak, since $O_{31}\sim 1$ but $O_{11}\sim
1/\sqrt{2}(m_e/m_\mu)$. Hence, we may neglect it when comparing
(\ref{tan2}) with the data on neutrino mixings.

The dependence of $\tan \theta_{12}$ on the phase $\phi_{\nu}$ and the
physical masses of the neutrinos enters through the ratio of the
neutrino mass differences under the square root sign, it can be made
explicit with the help of the unitarity constraint on $U_{\nu}$, 
eq. (\ref{phase-condition}),
%{\scriptsize
\be
%\begin{array}{l}
\frac{\displaystyle{m_{\nu_{2}}-m_{\nu_{3}}}}{
\displaystyle{m_{\nu_{3}}-m_{\nu_{1}}}}=%\\
%\\
\frac{(|m_{\nu_{2}}|^2-|m_{\nu_{3}}|^{2}\sin^{2}\phi_{\nu})^{1/2}-|m_{\nu_{3}}|
  |\cos
    \phi_{\nu}|}
{(|m_{\nu_{1}}|^{2}-|m_{\nu_{3}}|^{2}\sin^{2}\phi_{\nu})^{1/2}+|m_{\nu_{3}}|
  |\cos
    \phi_{\nu}|}.
%\end{array}
\label{tansq}
\ee
Similarly, the Majorana phases are given by
\be
\begin{split}
\sin 2\alpha=\sin(\phi_{1}-\phi_{2})=
\frac{|m_{\nu_{3}}|\sin\phi_{\nu}}{|m_{\nu_{1}}||m_{\nu_{2}}|}\times%\\
\\
\left(\sqrt{|m_{\nu_{2}}|^2-|m_{\nu_{3}}|^{2}\sin^{2}\phi_{\nu}}+\sqrt{|m_{\nu_{1}}|^{2}-|m_{\nu_{3}}|^{2}\sin^{2}\phi_{\nu}}\right),
\end{split}
\ee
\be
\begin{split}
\sin 2\beta=\sin(\phi_{1}-\phi_{\nu})=%\\
\\
 \frac{\sin\phi_{\nu}}{|m_{\nu_{1}}|}\left(|m_{\nu_{3}}|\sqrt{1-\sin^{2}\phi_{\nu}}+\sqrt{|m_{\nu_{1}}|^{2}-|m_{\nu_{3}}|^{2}\sin^{2}\phi_{\nu}}\right).
\end{split}
\ee
%}
A more complete and detailed discussion of the Majorana phases in the
neutrino mixing matrix $V_{PMNS}$ obtained in our model is given by 
J. Kubo~\cite{kubo-u}.

\begin{center}
{\it Neutrino masses and mixings}
\end{center}

In the present model, $\sin^{2} \theta_{13}$ and $\sin^{2} \theta_{23}$ are
determined by the masses of the charged leptons in very good
agreement with the experimental values~\cite{Maltoni:2004ei,schwetz,fogli1},
\be
\begin{array}{ll}
(\sin^{2}\theta_{13})^{th}=1.1\times 10^{-5}, &(\sin^2
  \theta_{13})^{exp} \leq 0.046, \nonumber
\end{array}
\ee
and
\be
\begin{array}{ll}
(\sin^{2}\theta_{23})^{th}=0.499, &(\sin^2
  \theta_{23})^{exp}=0.5^{+0.06}_{-0.05}.\nonumber
\end{array}
\ee
In this model, the experimental restriction $|\Delta
m^2_{12}|<|\Delta m^2_{13}|$ implies an inverted neutrino mass
spectrum, $|m_{\nu_{3}}|<|m_{\nu_{1}}|<|m_{\nu_{2}}|$~\cite{kubo1}.

As can be seen from eqs. (\ref{tan2}) and (\ref{tansq}), the solar mixing angle is
sensitive to the neutrino mass differences and the phase $\phi_{\nu}$
but is only very weakly sensitive to the charged lepton
masses. If we neglect the small terms proportional to $O_{11}$ and $O_{11}^2$
in (\ref{tan2}), we get
\be
\begin{array}{l}
\tan^2 \theta_{12} =
\frac{(\Delta m_{12}^2+\Delta m_{13}^2+|m_{\nu_{3}}|^{2}\cos^{2}\phi_{\nu})^{1/2}-|m_{\nu_{3}}|
  |\cos
    \phi_{\nu}|}
{(\Delta m_{13}^2+|m_{\nu_{3}}|^{2}\cos^{2}\phi_{\nu})^{1/2}+|m_{\nu_{3}}|
  |\cos
    \phi_{\nu}|}.
\end{array}
\label{tansq2}
\ee

\begin{figure}
\includegraphics[angle=270,width=15cm]{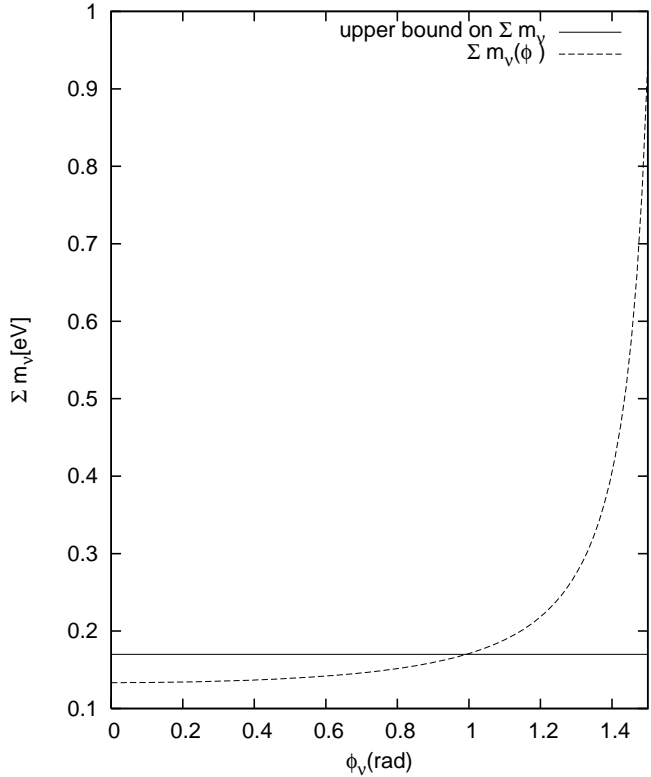}
\caption{\label{fig:epsart}The dashed line represents the sum of the
  neutrino masses, $\sum_{i=1}^{3}|m_{\nu_i}|$, as function of
  $\phi_\nu$. The horizontal straight line is the cosmological upper
  bound on $\sum |m_{\nu_i}|$~\cite{Seljak}.}
\end{figure}

From this expression, we may readily derive expressions for the
neutrino masses in terms of $\tan \theta_{12}$ and $\phi_\nu$ and the
differences of the squared masses
\be
|m_{\nu_{3}}|=\frac{\sqrt{\Delta m_{13}^2}}{2\cos \phi_{\nu} \tan
  \theta_{12}}\frac{1-\tan^4 \theta_{12}+r^2}{\sqrt{1+\tan^2 \theta_{12}} \sqrt{1+\tan^2 \theta_{12}+r^2}},
\label{masa3}
\ee
in a similar way, we obtain
\begin{widetext}
\be
\ba{l}
|m_{\nu_{1}}|=\frac{\sqrt{\Delta m_{13}^2}}{2\cos \phi_{\nu} t_{12}}\frac{
\left[(1-t_{12}^4)^2+4\cos^2
  \phi_{\nu}t_{12}^2(1+t_{12}^2)^2+2r^2\left(1-t_{12}^4+2\cos^2
    \phi_{\nu}t_{12}^2(1+t_{12}^2)\right)+r^4\right]^{(1/2)}}{\sqrt{1+t_{12}^2} \sqrt{1+t_{12}^2+r^2}}\\ \\
|m_{\nu_{2}}|=\frac{\sqrt{\Delta m_{13}^2}}{2\cos \phi_{\nu} t_{12}}\frac{
\left[(1-t_{12}^4)^2+4\cos^2 \phi_{\nu}t_{12}^2(1+t_{12}^2)^2+2r^2\left(1-t_{12}^4+2\cos^2 \phi_{\nu}t_{12}^2(1+t_{12}^2)(2+t_{12}^2)\right)+r^4\left(1+4\cos^2 \phi_{\nu}t_{12}^2(1+t_{12}^2)\right)\right]^{(1/2)}}{\sqrt{1+t_{12}^2} \sqrt{1+t_{12}^2+r^2}},\label{masa2}
\ea
\ee
\end{widetext}
where $t_{12}=\tan \theta_{12}$, and $r^2=\Delta m_{12}^2/\Delta
m_{13}^2\approx 3\times 10^{-2}$. As $r^2<<1$, eq. (\ref{masa3}) reduces to
\be
|m_{\nu_{3}}|\approx \frac{1}{2\cos \phi_{\nu}}\frac{\sqrt{\Delta
  m_{13}^2}}{ \tan \theta_{12}}(1-\tan^2 \theta_{12}).
\ee
From these expressions, and setting $r^2 \sim 0$, the sum of the
neutrino masses is
\begin{widetext}
\be
|m_{\nu_{1}}|+|m_{\nu_{2}}|+|m_{\nu_{3}}|\approx\frac{\sqrt{\Delta m_{13}^2}}{2\cos \phi_{\nu}
  \tan \theta_{12}}\left(1+2\sqrt{1+2\tan^2 \theta_{12}\left(2 \cos^2 \phi_\nu-1\right)+\tan^4 \theta_{12}}-\tan^2 \theta_{12}\right).
\ee
\end{widetext}
The most restrictive cosmological upper bound for this sum is~\cite{Seljak}
\be
\sum |m_{\nu}|\leq 0.17 eV.
\ee
From this upper bound and the experimentally determined values of
$\tan \theta_{12}$ and $\Delta m_{ij}^{2}$, we may derive a lower
bound for $\cos \phi_{\nu}$
\be
\cos \phi_{\nu}\geq 0.55
\ee
or $0\leq \phi_\nu \leq 57^{\circ}$. The neutrino masses $|m_{\nu_i}|$
assume their minimal values when $\cos \phi_\nu=1$. When $\cos
\phi_\nu$ takes values in the range $0.55 \leq \cos \phi \leq 1$, the
neutrino masses change very slowly with $\cos \phi_\nu$, see Figure ~\ref{fig:epsart}. In the
absence of experimental information we will assume that $\phi_\nu$ vanishes.
Hence, setting $\phi_\nu=0$ in our formula, we find
\be
\begin{split}
|m_{\nu_{2}}|\approx0.056eV,\\ |m_{\nu_{1}}|\approx 0.055eV,\\
|m_{\nu_{3}}|\approx 0.022eV,
\end{split}
\label{massesnum}
\ee
where we used the values $\Delta m^{2}_{13}=2.6 \times 10^{-3}eV^{2}$,
$\Delta m^{2}_{21}= 7.9 \times 10^{-5}eV^{2}$
and $\tan \theta_{12}=0.667$, taken
from ~\cite{Gonzalez-Garcia:2007ib}.

\begin{center}
{\it $V_{PMNS}$ and the tri-bimaximal form}
\end{center}

Once the numerical values of the neutrino masses are determined, we
may readily verify that the theoretical mixing matrix, $V_{PMNS}$, is
very close to the tri-bimaximal form of the mixing matrix,
\be
V_{PMNS}^{th}=
\left(\ba{ccc}
\sqrt{\frac{2}{3}} &\sqrt{\frac{1}{3}} & 0\\
-\sqrt{\frac{1}{6}} &\sqrt{\frac{1}{3}} & -\sqrt{\frac{1}{2}}\\
-\sqrt{\frac{1}{6}} &\sqrt{\frac{1}{3}} & \sqrt{\frac{1}{2}}
\ea\right)+\delta V_{PMNS}^{tri}
\ee
where $\delta V_{PMNS}^{tri}=V_{PMNS}^{th}-V_{PMNS}^{tri}$. From
eqs.(\ref{unitary-leptons-2}), (\ref{ununew}), (\ref{cosysin}),(\ref{vpmns2}), (\ref{mixing-angles}) and (\ref{massesnum}), the correction term
to the tri-bimaximal form of the mixing matrix comes out as
\be
\delta V_{PMNS}^{tri}\approx 
\left(\ba{ccc}
1.94 \times 10^{-2} & -2.84 \times 10^{-2}& -3.4\times 10^{-3}\\
2.21\times 10^{-2} & 1.5\times 10^{-2}& -8.2\times 10^{-6}\\
1.8\times 10^{-2} & 1.24\times 10^{-2}& 3.1\times 10^{-10}
\ea\right).
\ee
\section{Flavour Changing Neutral Currents (FCNC)}

Models with more than one Higgs $SU(2)$ doublet have tree level
flavour changing neutral currents. In the Minimal
$S_{3}$-invariant Extension of the Standard Model considered here,
there is one Higgs $SU(2)$ doublet per generation coupling to all fermions. The
flavour changing Yukawa couplings may be written in a flavour labelled,
symmetry adapted weak basis as
\be
\ba{lcl}
%\begin{align}
\hspace{-8pt}{\cal L}^{\rm FCNC}_{Y} =
\left(\overline{E}_{aL} Y_{a b}^{ES} E_{bR}+
\overline{U}_{aL} Y_{a b}^{US} U_{bR}
+\overline{D}_{aL} Y_{a b}^{DS} D_{bR}\right)H_S^0 \\ \\
~+\left(\overline{E}_{aL} Y_{a b}^{E1} E_{bR}+
\overline{U}_{aL} Y_{a b}^{U1} U_{bR}
+\overline{D}_{aL} Y_{a b}^{D1} D_{bR}\right)H_1^0+ \\ \\
\left(\overline{E}_{aL} Y_{a b}^{E2} E_{bR}+
\overline{U}_{aL} Y_{a b}^{U2} U_{bR}
+\overline{D}_{aL} Y_{a b}^{D2} D_{bR}\right)H_2^0+\mbox{h.c.}
\ea
\label{fcnf-lept}
%\end{align}
\ee
where the entries in the column matrices $E's$, $U's$ and $D's$ are
the left and right fermion fields and $Y_{ab}^{(e,u,d)s}$,
$Y_{ab}^{(e,u,d)1,2}$ are $3\times 3$ matrices of the Yukawa couplings
of the fermion fields to the neutral Higgs fields $H_{s}^0$ and
$H_{I}^0$ in the the $S_3$-singlet and doublet representations, respectively.

In this basis, the Yukawa couplings of the Higgs fields to each
family of fermions may be written in terms of matrices
${\cal{M}}_{Y}^{(e,u,d)}$, which give rise to the corresponding mass
matrices $M^{(e,u,d)}$ when the gauge symmetry
is spontaneously broken. From this relation we may calculate the
flavour changing Yukawa couplings in terms of the fermion masses and
the vacuum expectation values of the neutral Higgs fields. For
example, the matrix ${\cal{M}}_{Y}^e$ is written in terms of the
matrices of the Yukawa couplings of the charged leptons as
\be
{\cal{M}}_{Y}^e=Y_{w}^{E1} H^0_{1}+Y_{w}^{E2} H^0_{2},
\ee
in this expression, the index $w$ means that the Yukawa matrices are
defined in the weak basis,
\be
Y_{w}^{E1}=\frac{m_{\tau} }{v_1}\left(
\begin{array}{ccc}
0 & \frac{1}{\sqrt{2}}\frac{\tilde{m}_{\mu}}{\sqrt{1+x^2}} & \frac{1}{\sqrt{2}} \sqrt{\frac{1+x^2-\tilde{m}_{\mu}^2}{1+x^2}} \\ \\
 \frac{1}{\sqrt{2}}\frac{\tilde{m}_{\mu}}{\sqrt{1+x^2}} & 0  & 0 \\ \\
\frac{\tilde{m}_{e}(1+x^2)}{\sqrt{1+x^2-\tilde{m}_{\mu}^2}}e^{i\delta_{e}} & 0 & 0
\end{array}
\right)
\ee
and
\be
Y_{w}^{E2}=\frac{m_{\tau} }{v_2}\left( 
\begin{array}{ccc}
\frac{1}{\sqrt{2}}\frac{\tilde{m}_{\mu}}{\sqrt{1+x^2}} & 0 & 0 \\ \\
 0 &-\frac{1}{\sqrt{2}}\frac{\tilde{m}_{\mu}}{\sqrt{1+x^2}}  & \frac{1}{\sqrt{2}} \sqrt{\frac{1+x^2-\tilde{m}_{\mu}^2}{1+x^2}} \\ \\
0 & \frac{\tilde{m}_{e}(1+x^2)}{\sqrt{1+x^2-\tilde{m}_{\mu}^2}}e^{i\delta_{e}} & 0
\end{array}
\right).
\ee
The Yukawa couplings of immediate physical interest in the computation
of the flavour changing neutral currents are those defined in the mass
basis, according to
$\tilde{Y}_{m}^{EI}=U_{eL}^{\dagger}Y_{w}^{EI}U_{eR}$, where  $U_{eL}$ and
$U_{eR}$ are the matrices that diagonalize the charged lepton mass
matrix defined in eqs. (\ref{unu}) and (\ref{unitary-leptons}). We obtain 
\be
\tilde{Y}_{m}^{E1}\approx \frac{m_{\tau}}{v_{1}}\left(
\ba{ccc}
2\tilde{m}_{e} & -\frac{1}{2}\tilde{m}_{e} & \frac{1}{2} x \\
\\
-\tilde{m}_{\mu} & \frac{1}{2}\tilde{m}_{\mu} & -\frac{1}{2} \\
\\
\frac{1}{2} \tilde{m}_{\mu} x^2 & -\frac{1}{2}\tilde{m}_{\mu} & \frac{1}{2}
\ea
\right)_{m},
\label{y1m}
\ee
and
\be
\tilde{Y}_m^{E2}\approx \frac{m_{\tau}}{v_{2}}\left(
\ba{ccc}
-\tilde{m}_{e} & \frac{1}{2}\tilde{m}_{e} & -\frac{1}{2} x \\
\\
\tilde{m}_{\mu} & \frac{1}{2}\tilde{m}_{\mu} & \frac{1}{2} \\
\\
-\frac{1}{2} \tilde{m}_{\mu} x^2 & \frac{1}{2}\tilde{m}_{\mu} & \frac{1}{2}
\ea
\right)_{m},
\label{y2m}
\ee
where $\tilde{m}_{\mu}=5.94\times 10^{-2}$, $\tilde{m}_{e}=2.876 \times
10^{-4}$ and $x=m_{e}/m_{\mu}=4.84 \times 10^{-3}$.
All the non-diagonal elements are responsible for tree-level FCNC
processes. The actual values of the Yukawa couplings in
eqs. (\ref{y1m}) and (\ref{y2m}) still depend on the VEV's of the
Higgs fields $v_{1}$ and $v_{2}$, and, hence, on the Higgs
potential. If the $S_{2}^{\prime}$ symmetry in the Higgs sector is
preserved~\cite{pakvasa1}, $\langle H_{1}^{0}
\rangle = \langle H_{2}^{0} \rangle= v $. To make an order of magnitude estimate of the coefficient in the Yukawa
matrices, $m_\tau/v$, we may further assume that the VEV's
for all the Higgs fields are comparable, that is, $\langle H_{s}^{0} \rangle=\langle
H_{1}^{0}\rangle = \langle H_{2}^{0}
\rangle=\frac{\sqrt{2}}{\sqrt{3}}\frac{M_W}{g_2}$, then,  $m_\tau/v=\sqrt{3}/\sqrt{2}g_2m_\tau/M_W$
and we may estimate the numerical values of the Yukawa couplings from
the numerical %\ru\columnbreak
\noindent values of the lepton masses. For instance, the amplitude of the
flavour violating process $\tau^-\to \mu^-e^+e^-$, is
proportional to $\tilde{Y}_{\tau \mu}^{E}\tilde{Y}_{e e}^{E}$~\cite{Sher:1991km}. Then,
the leptonic branching ratio,
\be
Br(\tau \to \mu e^+ e^-)=\frac{\Gamma(\tau \to \mu e^+
  e^-)}{\Gamma(\tau \to e \nu \bar{\nu})+\Gamma(\tau \to \mu \nu \bar{\nu})}
\ee
and 
\be
\Gamma(\tau \to \mu e^+
  e^-)\approx \frac{m_{\tau}^5}{3\times 2^{10} \pi^3}\frac{\left(Y^{1,2}_{ \tau
      \mu}Y^{1,2}_{ e e}\right)^2}{M_{H_{1,2}}^4}
\ee
which is the dominant term, and the well known expressions for
$\Gamma(\tau \to e \nu \bar{\nu})$ and $\Gamma(\tau
\to \mu \nu \bar{\nu})$~\cite{PDG}, give
\be
Br(\tau \to \mu e^+
e^-)\approx\frac{9}{4}\left(\frac{m_{e}m_{\mu}}{m_{\tau}^2}\right)^2
\left(\frac{m_{\tau}}{M_{H_{1,2}}}\right)^4,
\ee
taking for $M_{H_{1,2}}\sim 120~GeV$, we obtain $$Br(\tau
\to \mu e^+ e^-)\approx 3.15 \times 10^{-17}$$ well below the
experimental upper bound for this process, which is $2.7 \times
10^{-7}$~\cite{aubert}. 
\begin{table*}
\caption{\label{table2}Leptonic FCNC processes, calculated with $M_{H_{1,2}}\sim 120~GeV$.}
\begin{ruledtabular}
\begin{tabular}{llll}
%\hline
FCNC processes & Theoretical BR &  Experimental  & References \\
& & upper bound BR &
\\ %\hline
$\tau \to  3\mu$ & $8.43
\times 10^{-14}$& $ 2 \times 10^{-7}$ & B. Aubert {\it et. al.} ~\cite{aubert}
 
\\ %\hline
$\tau \to  \mu e^+ e^-$ & $3.15 \times 10^{-17}$& $2.7 \times 10^{-7} $ &B. Aubert {\it et. al.} ~\cite{aubert}
 
\\ %\hline

$\tau \to \mu \gamma$ &  $9.24 \times 10^{-15}$ & $ 6.8 \times 10^{-8}$& B. Aubert {\it et. al.} ~\cite{aubert2} 
%\\ \hline
%$\mu \to  e \gamma < 1.2 \times 10^{-11}$ & PRL 83 (1999) 1521
\\ %\hline
$\tau \to e \gamma$ & $5.22\times 10^{-16}$ & $ 1.1 \times 10^{-11}$ &  B. Aubert {\it et. al.} ~\cite{aubert3}  
\\ %\hline 
$\mu \to  3e$ &  $2.53 \times 10^{-16}$ & $  1 \times 10^{-12}$ &
U. Bellgardt {\it et al.} ~\cite{bellgardt}  
\\ %\hline
$\mu \to e \gamma$ &  $2.42 \times 10^{-20}$ & $ 1.2 \times 10^{-11}$& M.~L.~Brooks {\it et al.} ~\cite{Brooks:1999pu}
\end{tabular}
\end{ruledtabular}
\end{table*}
Similar computations give the following estimates
\be
%\ba{lcl}
Br(\tau \to e
\gamma)\approx
\frac{3\alpha}{8\pi}\left(\frac{m_\mu}{M_H}\right)^4,
\ee
\be
%\ba{lcl}
Br(\tau \to \mu
\gamma)\approx\frac{3\alpha}{128\pi}\left(\frac{m_{\mu}}{m_{\tau}}\right)^2\left(\frac{m_\tau}{M_H}\right)^4,
\ee
\be
Br(\tau \to 3\mu)\approx\frac{9}{64}\left(\frac{m_\mu}{M_H}\right)^4,
\ee
\be
Br(\mu \to 3e)\approx 18 \left(\frac{m_e m_\mu}{m_\tau^2}\right)^2\left(\frac{m_\tau}{M_H}\right)^4,
\ee
and 
\be
%\ba{lcl}
%\begin{split}
Br(\mu \to e
\gamma)\approx
\frac{27\alpha}{64\pi}\left(\frac{m_e}{m_\mu}\right)^4\left(\frac{m_\tau}{M_H}\right)^4.
%\ea
%\end{split}
\ee
We see that FCNC processes in the leptonic sector are strongly
suppressed by the small values of the mass ratios
$m_e/m_\tau$, $m_\mu/m_\tau$ and
$m_\tau/M_H$. The numerical estimates of the branching
ratios and the corresponding experimental upper bounds are shown in
Table~\ref{table2}. It may be seen that, in all cases considered, the numerical
values for the branching ratios of the FCNC in the leptonic sector are
well below the corresponding experimental upper bounds. 
%\columnbreak\rd \noindent
The matrices of the quark Yukawa couplings may be computed in a
similar way. Numerical values for the Yukawa couplings for u and
d-type quarks are given in our previous paper ~\cite{kubo1}. There, it was
found that, due to the strong hierarchy in the quark masses and the
corresponding small or very small
 mass ratios, the numerical values of
all the Yukawa couplings in the quark sector are small or very
small. Kubo, Okada and Sakamaki~\cite{kubo-pot} have investigated the breaking of the
gauge symmetry in the present $S_3$-invariant extension of the
Standard Model with the $S_3$-invariant Higgs potential
$V_H(H_S,H_{{\bf 2}})$ analyzed by Pakvasa and Sugawara ~\cite{pakvasa1}. They found that it is possible that all physical Higgs
bosons, except one neutral one, could become sufficiently heavy
($M_H\sim 10~TeV$) to suppress all the flavour changing neutral
current processes in the quark sector of the theory without having a
problem with triviality.

\section{Conclusions}
By introducing three Higgs fields that are $SU(2)_{L}$ doublets in the
theory, we extended the concept of flavour and generations to the
Higgs sector and formulated a Minimal $S_{3}$-Invariant Extension of
the Standard Model~\cite{kubo1}. A well defined structure of the Yukawa
couplings is obtained, which permits the calculation of mass and mixing
matrices for quarks and leptons in a unified way. A further reduction
of redundant parameters is achieved in the leptonic sector by
introducing a $Z_{2}$ symmetry. The flavour symmetry group $Z_{2}
\times S_{3}$ relates the mass spectrum and mixings. This allowed us
to derive explicit, analytical expressions for all entries in the
neutrino mixing matrix, $V_{PMNS}$ as functions of the masses of the
charged leptons and neutrinos and two phases $\delta$ and
$\phi_{\nu}$~\cite{Felix:2006pn}. In this model, the tri-bimaximal mixing
structure of $V_{PMNS}$ and the magnitudes of the three mixing angles
are determined by the interplay of the flavour $S_{3}\times Z_{2}$
symmetry, the see-saw mechanism and the charged lepton mass
hierarchy. We also found that $V_{PMNS}$ has three CP violating
phases, namely, one Dirac phase $\delta=\delta_{\nu}-\delta_{e}$ and two 
Majorana phases, $\alpha$ and $\beta$, which are functions of the
neutrino masses and the phase $\phi_{\nu}$ which is independent of the
Dirac phase. The numerical
values of the reactor, $\theta_{13}$, and the atmospheric,
$\theta_{23}$, mixing angles are determined by the masses of the
charged leptons only, in very good agreement with the experiment. The
solar mixing angle $\theta_{12}$ is almost insensitive to the values
of the masses of the charged leptons, but its experimental value
allowed us to fix the scale and origin of the neutrino mass spectrum,
which has an inverted hierarchy, with the values $|m_{\nu_{2}}|=0.056eV$,
$|m_{\nu_{1}}|=0.055eV$ and $|m_{\nu_{3}}|=0.022eV$.
In the present work, we obtained explicit expressions for the matrices of the Yukawa
couplings of the lepton sector parametrized in terms of the charged lepton masses and the
VEV's of the neutral Higgs bosons in the $S_3$-doublet
representation. These Yukawa matrices are closely related to the
fermion mass matrices and have a structure of small and very small
entries reflecting the observed charged lepton mass hierarchy. With
the help of the Yukawa matrices, we computed the branching ratios of a
number of FCNC processes and found that the branching ratios of all
FCNC processes considered are strongly suppressed by powers of the
small mass ratios $m_e/m_\tau$  and $m_\mu/m_\tau$, and by the ratio
$\left(m_\tau/M_{H_{1,2}}\right)^4$, where $M_{H_{1,2}}$ is the mass
of the neutral Higgs bosons in the $S_3$-doublet. Taking for
$M_{H_{1,2}}$  a very conservative value ($M_{H_{1,2}}\approx
120~GeV$), we found that the numerical values of the branching ratios
of the FCNC in the leptonic sector are well below the corresponding
experimental upper bounds by many orders of magnitude.
We may add that although the theoretical values of the branching
ratios of FCNC processes computed in this work are much smaller than
their experimental upper bounds measured in terrestrial laboratories,
they still are larger than the vanishing or nearly vanishing
values allowed by the Standard Model, and could be important in
astrophysical processes~\cite{Raffelt:2007nv}. It has already been argued that small FCNC
processes mediating non-standard quark-neutrino interactions could be
important in the theoretical description of the gravitational core
collapse and shock generation in the explosion stage of a supernova~\cite{Amanik:2004vm,valle-nu}.
\section{Acknowledgements}
We are indebted to Prof. S. Pakvasa for useful comments on this
work. This work was partially supported by CONACYT M\'exico under contract
No 51554-F and by DGAPA-UNAM under contract PAPIIT-IN115207-2.


\begin{thebibliography}{9}
 
\bibitem{jung}C. K. Jung, C. Mc Grew, T. Kajita and T. Mann,
    {\it Annu. Rev. Nucl. Part. Sci}, {\bf 51} (2001) 451.
\bibitem{Mohapatra:2006gs}
  R.~N.~Mohapatra and A.~Y.~Smirnov,
  %``Neutrino mass and new physics,''
 {\it Ann.\ Rev.\ Nucl.\ Part.\ Sci.\ } {\bf 56}, 569
 (2006).arXiv:hep-ph/0603118.
\bibitem{altmann} M.~Altmann  {\em et al.} [GNO collaboration], {\it
  Phys. Lett.} B {\bf 616}, (2005), 174.

\bibitem{smy} M. B. Smy {\em et al.} [SK collaboration], {\it
  Phys. Rev.} D {\bf 69} (2004), 011104.

\bibitem{ahmad} Q.R.~Ahmad {\em et al.} [SNO collaboration], {\it
    Phys. Rev. Lett}. {\bf 89} (2002) 011301.

\bibitem{aharmim} B. Aharmim {\em et al.} [SNO collaboration], {\it
    Phys. Rev.} {\bf C72} (2005), 055502.  \mbox{arxiv: nucl-ex/0502021.}
\bibitem{fukuda} S.~Fukuda {\it et al.} ,[SK collaboration]
  {\it Phys. Lett}. {\bf B539} (2002) 179.
\bibitem{Ashie:2005ik}
  Y.~Ashie {\it et al.},
  %[Super-Kamiokande Collaboration],
  %``A measurement of atmospheric neutrino oscillation parameters by
  %Super-Kamiokande I,''
 {\it  Phys. Rev. Lett.} {\bf 93} (2004) 101801; [hep-ex/0404034]. 
\bibitem{bemporad}  C. Bemporad, G. Gratta and P. Vogel, {\it
  Rev. Mod. Phys.}, {\bf 74} (2002) 297.
\bibitem{Araki:2004mb}
  T.~Araki {\it et al.}  (KamLAND collaboration),
  %``Measurement of neutrino oscillation with KamLAND: Evidence of spectral
  %distortion,''
{\it  Phys. Rev. Lett.}  {\bf 94}, (2005) 081801.
\bibitem{Maltoni:2004ei}
  M.~Maltoni, T.~Schwetz, M.A.~T{\'o}rtola and J.W.F.~Valle,
  %``Status of global fits to neutrino oscillations,''
{\it   New J. Phys}.  {\bf 6} (2004) 122.
\bibitem{schwetz} T.~Schwetz, ``Neutrino oscillations: Current status
  and prospects'', {\it Acta Phys. Polon}. {\bf B36} (2005) 3203.
 arxiv: hep-ph/0510331.
\bibitem{Gonzalez-Garcia:2007ib}
  M.~C.~Gonz\'alez-Garcia and M.~Maltoni,
  %``Phenomenology with massive neutrinos,''
  arXiv:hep-ph/0704.1800.
  %%CITATION = ARXIV:0704.1800;%%
\bibitem{chooz} M. Apollonio et al. [CHOOZ Collaboration], {\it
  Eur. Phys. J.} {\bf C27} (2003)331 .
\bibitem{eitel} K. Eitel in ``Neutrino 2004'', 21st International
Conference on Neutrino Physics and Astrophysics (Paris, France 2004)
Ed. J. Dumarchey, Th. Patyak and F. Vanuucci. {\it Nucl. Phys.} B
({\it Proc Suppl.}) {\bf 143}, (2005) 197.

\bibitem{eliot} S. R. Eliot and J. Engel, {\it J. Phys.}  {\bf G 30}
(2004) R183.

\bibitem{Seljak}
  U.~Seljak, A.~Slosar and P.~McDonald,
  %``Cosmological parameters from combining the Lyman-alpha forest with CMB,
  %galaxy clustering and SN constraints,''
  {\it JCAP} {\bf 0610}, (2006) 014. arXiv:astro-ph/0604335.

\bibitem{elgaroy}  O. Elgaroy, and O. Lahav, {\it  New J. Phys.} {\bf
7} (2005) 61.

\bibitem{Lesgourgues}
  J.~Lesgourgues and S.~Pastor,
  %``Massive neutrinos and cosmology,''
  {\it Phys.\ Rept.}  {\bf 429}, (2006) 307. arXiv:astro-ph/0603494.
  %%CITATION = PRPLC,429,307;%%

\bibitem{kubo1} J. Kubo, A. Mondrag\'on, M. Mondrag\'on, E. Rodr\'{\i}guez-J\'auregui, {\it Prog. Theor. Phys.}  {\bf 109}, (2003), 795.

\bibitem{Felix:2006pn}  O.~F\'elix, A.~Mondrag\'on, M.~Mondrag\'on and
  E.~Peinado, {\it Rev. Mex. F\'{\i}s.}  {\bf S 52}, (2006),
  67-73. \mbox{arxiv: hep-ph/0610061.}
\bibitem{Fritzsc1} H.~Fritzsch  {\it Phys. Lett}. {\bf B70}, (1977), 436.
\bibitem{pakvasa1} S.~Pakvasa and H.~Sugawara, 
{\it Phys. Lett}. {\bf 73B} (1978), 61.
\bibitem{Fritzsc2} H.~Fritzsch,  {\it Phys. Lett}. {\bf B73}, (1978), 317.
\bibitem{harari}H.Harari, H.Haut, J.Weyers, {\it Phys. Lett}. {\bf B78}
	(1978), 459.
\bibitem{Frere:1978ds}
  J.~M.~Frere,  %``On The Use Of Permutation Symmetry,''
  {\it Phys.\ Lett.}   {\bf B80} (1979) 369.
\bibitem{Fritzsc3}H.~Fritzsch,  {\it Nucl. Phys.} B {\bf 155}, (1979), 189.
\bibitem{yamanaka}Y.~Yamanaka, S.~Pakvasa and H.~Sugawara, {\it Phys. Rev.} {\bf D25} (1982), 1895. Erratum-ibid.\  {\bf D 29}, 2135 (1984).
\bibitem{kaus}P.~Kaus and S.~Meshkov, {\it Phys. Rev.} {\bf D42} (1990), 1863.
\bibitem{Fritzsch4}H.~Fritzsch and J.P.~Plankl, {\it Phys. Lett.} {\bf B237} (1990), 451.
\bibitem{Harrison}P.F.~Harrison and W.G.~Scott, {\it Phys. Lett.} {\bf B333} (1994), 471.
\bibitem{mondragon1}A.~Mondrag{\'o}n  and
E.~Rodr\' iguez-J\' auregui, {\it Phys. Rev.} {\bf D59} (1999), 093009.
\bibitem{mondragon2}A.~Mondrag{\'o}n  and
E.~Rodr\' iguez-J\' auregui, {\it Phys. Rev.} {\bf D61} (2000), 113002.
\bibitem{xing} For a review see H.~Fritzsch and Z.Z.~Xing,
{\it Prog. Part. Nucl. Phys}. {\bf 45} (2000) 1.
\bibitem{hall}L.J.~Hall and H.~Murayama, 
{\it Phys. Rev. Lett}. {\bf 75} (1995),  3985.
\bibitem{hall2}C.D.~Carone, L.J.~Hall and H.~Murayama, 
{\it Phys. Rev}. {\bf D53} (1996), 6282.
\bibitem{koide} Y.~Koide, {\it  Phys. Rev}. {\bf D60} (1999), 077301.
\bibitem{ma}E.~Ma,  {\it Phys. Rev.} {\bf D44} (1991), 587.
\bibitem{ma2}E.~Ma, {\it  Mod. Phys. Lett}. {\bf A17} (2002), 627; ibid {\bf A17} (2002), 2361.
\bibitem{babu}K.S.~Babu, E.~Ma and J.W.F.~Valle, {\it Phys. Lett}. {\bf B552} (2003), 207.
\bibitem{chen}S.-L.Chen, M. Frigerio and E. Ma,  {\it Phys. Rev.}
{\bf D70}, 073008 (2004); Erratum: ibid {\bf D70} (2004), 079905.
\bibitem{grimus-la}W. Grimus and Lavoura {\it JHEP.} {\bf 0508}
 (2005) 013.
\bibitem{Hagedorn:2006ug}
  C.~Hagedorn, M.~Lindner and R.~N.~Mohapatra,
  %``S(4) flavor symmetry and fermion masses: Towards a grand unified theory  of
  %flavor,''
  {\it JHEP} {\bf 0606}, (2006) 042.
  arXiv:hep-ph/0602244.
  %%CITATION = JHEPA,0606,042;%%
\bibitem{Hagedorn:2006ir}
  C.~Hagedorn, M.~Lindner and F.~Plentinger,
  %``The discrete flavor symmetry D(5),''
  {\it Phys.\ Rev.}   {\bf D 74}, (2006) 025007.
  arXiv:hep-ph/0604265.
\bibitem{Haba:2005ds}
  N.~Haba and K.~Yoshioka,
  %``Discrete flavor symmetry, dynamical mass textures, and grand
  %unification,''
  {\it Nucl. Phys.}   {\bf B 739}, (2006) 254. 
arXiv:hep-ph/0511108.
  %%CITATION = NUPHA,B739,254;%%
\bibitem{Kaneko:2006wi}
  S.~Kaneko, H.~Sawanaka, T.~Shingai, M.~Tanimoto and K.~Yoshioka,
  %``Flavor symmetry and vacuum aligned mass textures,''
  {\it Prog. Theor. Phys.}  {\bf 117}, (2007) 161. 
  arXiv:hep-ph/0609220.
  %%CITATION = PTPKA,117,161;%%
\bibitem{Smirnov:2006qz}
  A.~Y.~Smirnov,
  %``Neutrino mass and new physics,''
  {\it J.\ Phys.\ Conf.\ Ser.}  {\bf 53}, (2006) 44.
\bibitem{Altarelli:2004za}
  G.~Altarelli and F.~Feruglio,
  %``Models of neutrino masses and mixings,''
  {\it New J.\ Phys.}  {\bf 6}, (2004) 106; G.~Altarelli, {\it In the Proceedings of IPM School and Conference on Lepton and Hadron Physics (IPM-LHP06), Tehran, Iran, 15-20 May 2006, pp 0001},  arXiv:hep-ph/0610164.

  %%CITATION = NJOPF,6,106;%%
\bibitem{Mondragon:2006hi}
  A.~Mondrag\'on, {\it AIP Conf Ser} {\bf 857B} (2006) 266-282.  %``Models of flavour with discrete symmetries,''
 arXiv:hep-ph/0609243.

\bibitem{Albright:2006cw}
  C.~H.~Albright and M.~C.~Chen,
  %``Model predictions for neutrino oscillation parameters,''
  {\it Phys.\ Rev.}   {\bf D74},  (2006) 113006. arXiv:hep-ph/0608137.

\bibitem{PDG} W-M Yao {\it et al.} [Particle Data Group],{\it J. Phys. G: Nucl. Part. Phys.}
  {\bf 33} (2006) 1.
\bibitem{Harrison:2002er}
  P.~F.~Harrison, D.~H.~Perkins and W.~G.~Scott,
  %``Tri-bimaximal mixing and the neutrino oscillation data,''
  {\it Phys. Lett.}   {\bf B530}, (2002) 167.
  arXiv:hep-ph/0202074.
  %%CITATION = PHLTA,B530,167;%%
\bibitem{kubo-u}J. Kubo,  {\it Phys. Lett.} {\bf B578}, (2004), 156; Erratum:
ibid {\bf 619} (2005) 387.
\bibitem{fogli1} G.~L.~Fogli, E.~Lisi, A.~Marrone and A.~Palazzo, {\it
    Prog. Part. Nucl. Phys.} {\bf 57} (2006) 742-795. 
  arxiv: hep-ph/0506083.
 \bibitem{Seljak:2006bg}
  U.~Seljak, A.~Slosar and P.~McDonald,
  %``Cosmological parameters from combining the Lyman-alpha forest with CMB,
  %galaxy clustering and SN constraints,''
  {\it JCAP} {\bf 0610}, (2006) 014. arXiv:astro-ph/0604335.
  %%CITATION = JCAPA,0610,014;%%
%\bibitem{serra}  G.~L.~Fogli, E.~Lisi, A.~Marrone, A.~Melchiorri,
%   A.~Palazzo, P.~Serra and J.~Silk, {\it  Phys. Rev.}  {\bf D70}
%   (2004), 113003. 
\bibitem{Sher:1991km}
  M.~Sher and Y.~Yuan,  %``Rare B decays, rare tau decays and grand unification,''
  {\it Phys.\ Rev.}   {\bf D44}, 1461 (1991).
  %%CITATION = PHRVA,D44,1461;%%
 \bibitem{aubert}  B. Aubert {\it et al.} [BABAR Collaboration],{\it
     Phys. Rev. Lett}. {\bf 92} (2004), 121801.
 \bibitem{aubert2} B. Aubert {\it et al.} [BABAR Collaboration], {\it
     Phys. Rev. Lett}. {\bf 95} (2005), 041802.
 \bibitem{aubert3} B. Aubert {\it et al.} [BABAR Collaboration], {\it Phys. Rev. Lett}. {\bf 96} (2006), 041801.
\bibitem{bellgardt} U. Bellgardt {\it et al.} [SINDRUM
  Collaboration], {\it Nucl. Phys.} {\bf B299} (1998) 1.
\bibitem{Brooks:1999pu}
  M.~L.~Brooks {\it et al.}  [MEGA Collaboration],
  %``New limit for the family-number non-conserving decay mu+ --> e+ gamma,''
  {\it Phys.\ Rev.\ Lett.}  {\bf 83}, (1999) 1521. arXiv:hep-ex/9905013.
  %%CITATION = PRLTA,83,1521;%%
 \bibitem{kubo-pot}  J. Kubo, H. Okada and F. Sakamaki, {\it Phys. Rev.} {\bf D70} (2004), 036007.
 %\bibitem{mondra}  A. Mondrag\'on, M. Mondrag\'on and E. Peinado, Work
 %  in progress.
\bibitem{Raffelt:2007nv}
  G.~G.~Raffelt,``Supernova neutrino observations: What can we
  learn?,'' To appear in the proceedings of 22nd International Conference on Neutrino Physics and Astrophysics (Neutrino 2006), Santa Fe, New Mexico, 13-19 Jun 2006. arXiv:astro-ph/0701677.

\bibitem{Amanik:2004vm}
  P.~S.~Amanik, G.~M.~Fuller and B.~Grinstein,
  %``Flavor changing supersymmetry interactions in a supernova,''
  {\it Astropart. Phys.}  {\bf 24}, (2005) 160. arxiv:hep-ph/0407130.
\bibitem{valle-nu}
A.~Esteban-Pretel, R.~Tomas and J.~W.~F.~Valle,  ``Probing
non-standard neutrino interactions with supernova neutrinos,'', arXiv:0704.0032 [hep-ph].
%  A. Esteban-Pretel, R. Tom\'{a}s and j. W. F. Valle, "Probing
%  non-standard neutrino interactions with supernova neutrinos", arxiv-hep:0704.0032.

\end{thebibliography}
\end{document}